# The Development and Performance of a Machine-Learning Based Mobile Platform for Visually Determining the Etiology of Penile Pathology

Lao-Tzu Allan-Blitz MD MPH[1]*, Sithira Ambepitiya MBBS[2], Raghavendra Tirupathi MD[3], Jeffrey D. Klausner MD MPH[4], Yudara Kularathne MD, FAMS[2]

(1). Division of Global Health Equity: Department of Medicine, Brigham and Women's Hospital, Boston, MA, USA

(2). HeHealth Inc. San Francisco, CA, USA

(3). Keystone Infectious Diseases, Keystone Health, Chambersburg, PA, USA

(4). Department of Population and Public Health Sciences, Keck School of Medicine, University of Southern California, Los Angeles, CA, USA

**Disclosures:** LAB and JDK received consulting fees from HeHealth Inc. SA is a Medical Executive at HeHealth Inc.

**Acknowledgements and Funding:** A startup, HeHealth has raised funding from institutional investors and angel investors and would like to specifically acknowledge Plug and Play Tech Center as well as ARKRAY Corporate Venture Capital.

**Corresponding Author Contact Information:**

Lao-Tzu Allan-Blitz

Department of Medicine, Brigham and Women's Hospital 75 Francis Street

Boston, MA 02115, USA

Email: lallan-blitz@partners.org, phone: (805) 896-5313

**Alternative Author Information:**

Prof. Jeffrey D. Klausner

Department of Population and Public Health Sciences Keck School of Medicine

University of Southern California Los Angeles, CA 90033, USA

Email: jdklausner@med.usc.edu

# Abstract

Machine-learning algorithms can facilitate low-cost, user-guided visual diagnostic platforms for addressing disparities in access to sexual health services. We developed a clinical image dataset using original and augmented images for five penile diseases: herpes eruption, syphilitic chancres, penile candidiasis, penile cancer, and genital warts. We used a U-net architecture model for semantic pixel segmentation into background or subject image, the Inception-ResNet version 2 neural architecture to classify each pixel as diseased or non-diseased, and a salience map using GradCAM++. We trained the model on a random 91% sample of the image database using 150 epochs per image, and evaluated the model on the remaining 9% of images, assessing recall (or sensitivity), precision, specificity, and F1-score (accuracy). Of the 239 images in the validation dataset, 45 (18.8%) were of genital warts, 43 (18.0%) were of HSV infection, 29 (12.1%) were of penile cancer, 40 (16.7%) were of penile candidiasis, 37 (15.5%) were of syphilitic chancres, and 45 (18.8%) were of non- diseased penises. The overall accuracy of the model for correctly classifying the diseased image was 0.944. Between July 1st and October 1st 2023, there were 2,640 unique users of the mobile platform. Among a random sample of submissions (n=437), 271 (62.0%) were from the United States, 64 (14.6%) from Singapore, 41 (9.4%) from Candia, 40 (9.2%) from the United Kingdom, and 21 (4.8%) from Vietnam. The majority (n=277 [63.4%]) were between 18 and 30 years old. We report on the development of a machine-learning model for classifying five penile diseases, which demonstrated excellent performance on a validation dataset. That model is currently in use globally and has the potential to improve access to diagnostic services for penile diseases.

**Keywords:** *Machine-Learning; Sexually Transmitted Infections; Visual Diagnostics; Mobile Technology; Health Equity*

# Research In Context

## Evidence before this study

Access to diagnostic services is a major barrier to sexual health care, and has numerous drivers, which include insufficient laboratory infrastructure in low-resource settings, lack of locally available health centers in rural areas, and stigma around sexually transmitted infections, which limits care-seeking. Machine-learning algorithms are increasingly being utilized in healthcare contexts, and can be used to visually categorize disease states.

## Added value of this study

We developed a machine-learning algorithm for classifying five penile pathologies based on user-submitted images via a mobile web-based application. We report on the methods used to develop that algorithm and the performance of that algorithm, demonstrating a high degree of accuracy for all five pathologies. We also report on the demographics of current users of the mobile web-based platform.

## Implications of all the available evidence

The machine-learning algorithm for classifying penile pathology has the potential to dramatically improve sexual health care via increasing access to diagnostic services. Future work may aim to expand the types of pathology recognized by the machine- learning algorithm and compare the performance of the algorithm with other forms of diagnostic testing.

# 1 Introduction

Sexually transmitted infections cause major disruptions to sexual, reproductive, and neonatal health, with a range of consequences among those with untreated infections including infertility[1] an increased risk for HIV transmission and acquisition,[2,3] cervical cancer,[4,5] neonatal blindness,[6] and fetal demise.[7] The rates of syphilis, caused by Treponema pallidum, have risen dramatically over the past several years.[8] The prevalence of human papilloma virus infection is estimated to be approximately 40% among individuals between the ages of 15 and 59 years in the United States,[9] and more than 30% globally,[10,11] Herpes simplex virus (HSV) 2, the leading cause of genital herpes, is estimated to affect 12% of individuals between 15 and 49 years old in the United States,[12] and 13% globally.[13]

Importantly, resource-limited settings often suffer disproportionate rates of sexually transmitted infections.[14] Further, such areas face numerous barriers to adequate diagnostics for sexually transmitted infections, including insufficient laboratory infrastructure to support microbiologic diagnoses, challenges in patients in rural settings accessing healthcare, and the heightened stigma around sexually transmitted infections, which limits care-seeking[15] and partner notification.[16] For many sexually transmitted diseases, however, cutaneous manifestations may be characteristic, such as herpes eruptions caused by HSV, genital warts caused by the human papilloma virus, chancers caused by Treponema pallidium, and penile candidiasis. Technologic solutions for improving access to visual diagnostics, therefore, have the potential to have a large impact on the burden of several sexually transmitted diseases and other penile disease states that can be diagnosed via visual inspection.

Artificial intelligence constitutes one such technologic solution. A sub-type of artificial intelligence, machine-learning, whereby a complex algorithm uses additional data to develop its own algorithms to solve a problem, has been used in an array of medical contexts, including sexual health care.[17] Artificial intelligence has been used to predict who might benefit most from pre-exposure prophylaxis to HIV,[18] the county-level rates of syphilis based on social media behaviors,[19] and an individual's risk for HIV and specific sexually transmitted infections based on response to routine surveys.[20] After the 2022 mpox (formerly monkey pox) outbreak, numerous machine-learning models were developed to visually diagnose mpox, which demonstrated diagnostic accuracies that ranged between 78-98%.[21] We have developed a machine-learning platform to diagnose five penile disease states: herpes eruption, syphilitic chancres, penile candidiasis, penile cancer, and genital warts. We present an overview of that platform and the technologic innovations involved in its development. We further highlight areas for future work exploring the potential utility of such a tool for addressing disparities in access to sexual health care.

# 2 Methods

Here were describe the development of a machine-learning algorithm. That development includes clinical image dataset generation, the development of models for semantic segmentation, classification, and salience modules, and finally training the algorithm and assessing performance on a subset of the image database.

## 2.1 Clinical Image Dataset Generation

We developed a repository of clinical images over two phases. Phase one consisted of obtaining clinical images from physicians of five penile diseases: herpes eruption, syphilitic chancres, penile candidiasis, penile cancer, and genital warts.

Clinicians who contributed images were specialists in infectious diseases, sexually transmitted diseases, dermatology, or family practice with a special interest in sexually transmitted diseases. Those clinicians practiced in six different countries: India (n=4), Sri Lanka (n=3), Singapore (n=3), Australia (n=3), the United States (n=1), and the United Kingdom (n=1).

In addition, we used a custom-built web-scraping tool to download images freely available on the internet. All images underwent expert evaluation, with labeling of distinguishable disease characteristics, and were ultimately categorized as one of the five penile diseases of interest. Experts who contributed to labeling and classifying the clinical images were from Indian (n=4), Sri Lanka (n=3), Singapore (n=3), Australia (n=3), the United States (n=1) and the United Kingdom (n=1). Phase one resulted in a total of 1,000 labeled and categorized images.

Phase two involved three components. First, we sought additional de-identified clinical images from clinicians in Singapore, India, and Sri Lanka. Second, we publicly sourced images from the mobile app interface, which was predominantly used in North America. Finally, we implemented layered image augmentation to offset the unequal distribution of clinical images across the five disease categories.

Layered image augmentation involves two stages: first, we conducted manual image augmentation by extracting specific visually recognizable disease patterns from the existing clinical image dataset and layering those patterns on top of images of health penises, accounting for disease location, skin complexation, and penis orientation. Those disease patterns were then overlay on top of a separate dataset of non-diseased penises to create additional, artificially modified images of the five penile diseases. For each manually augmented image, an expert clinician verified the validity of the image. Subsequently, we implemented automated random image augmentation, which randomly applies selected transformations to each image during the model training process to generate permutations of a given image. Those permutations included rotations, re-scaling of the image size, shifts in the x- and y-axis, changes in brightness, vertical or horizontal flips, and alterations in size, and color. We used Generative Adversarial Networks (GAN) technology to generate the modified images.[22] Supplemental Figure 1 in the Appendix shows an example of permutated images using various transformations that were included in the augmented image dataset for model training. Phase two resulted in a total of 2,627 images (30% contributed by clinicians, 30% sourced through the mobile app, and 40% developed via image augmentation) (Figure 1).

## 2.2 Model Development: Semantic Segmentation, Classification, and Salience Modules

We utilized semantic image segmentation, in which each pixel of a given image is classified to one of several pre-defined classes to isolate the visually distinguishable diseased regions for each of the five penile disease classifications, and to disregard the background and normal penis components. First, we used a U-net architecture model in order to generate a binary pixel classification as either background or subject image.[23] In order to further classify images, we applied the Inception-ResNet version 2 neural architecture,[24] a pre-trained model developed to classify images from a generic dataset and learn visual correlations. We used the Adam optimizer to assign weights to the images classified by the Inception-ResNet version 2 model in order to further focus on identification of images of a penis. The model was trained to conduct two predictions: the penile disease classification, and salience of the classification prediction. We mapped the predicted salience using GradCAM++, which facilitated visual explanations of model prediction results and localization of identified pathology (Supplemental Figure 2 in the Appendix).[25]

We further optimized the model using an image processing module in order to focus on diseased regions and reduce background signals. That was done by using the segmentation mask determined by the semantic segmentation module within a bounding box determined by the salient regions identified by the initial model. The pixels within the bounding box were then re-input as the image for which classification would be assessed using GradCAM++.

To refine the model, as well as assess model performance, we developed a training dataset by randomly selecting 91% of the overall images (both original and augmented images) using 150 iterations (epochs) per image. The learning rate and epsilon of the optimizer was set to 0.01 and 0.1, respectively. We trained the model with tesla p100 16gb graphics processing unit provide by Kaggle cloud platform (San Francisco, United States). We then evaluated the performance of the model on the remaining 9% of images (n=239). Evaluation metrics included: recall (or sensitivity), precision, specificity, and F1-score.[26] Equations for the outcome metrics are shown below. Overall accuracy of the model was defined as the F1-score averaged across each disease class.

**Equations**

$$Recall\ (Sensitivity) = \frac{True\ Positive}{(True\ Positive + False\ Negative)}$$

$$Precision = \frac{True\ Positive}{(True\ Positive + False\ Positive)}$$

# 3  Results

## 3.1  Performance of the Machine-Learning Algorithm

Of the 239 images in the validation dataset, 45 (18.8%) were of genital warts, 43 (18.0%) were of HSV infection, 29 (12.1%) were of penile cancer, 40 (16.7%) were of penile candidiasis, 37 (15.5%) were of syphilitic chancres, and 45 (18.8%) were of non- diseased penises. Table 1 shows the classification assignment and performance metrics of the model for each category. The overall accuracy of the model for correctly classifying the diseased image was 0.944.

## 3.2  Current Usage of the HeHealth Platform for Classifying Five Penile Diseases

As of October 1st 2023, there have been 37,100 unique submissions using the HeHealth platform. After the latest software on July 1st there were 2,640 unique submissions. We evaluated a random set of 437 submissions between July and October 2023. The majority of those users (n=271 [62.0%]) were in the United States, with 277 (63.4%) between the age of 18 to 30 years old. Table 2 shows the distribution of users by country, as well as symptoms reported, and recency of last sexual contact. Figure 2 shows examples of de-identified clinical images submitted by users during that time period.

# 4  Discussion

We report on the development of a machine-learning model for classifying five penile diseases. Based on more than 2,000 images (original clinical images and augmented images) from a geographically diverse sample, the model demonstrated excellent discrimination between for each of the five disease classifications and for non-diseased states. Use of this tool has the potential to transform the care of penile diseases, particularly in settings with limited access to care.

Model recall (or sensitivity) was high for genital warts, and herpes eruption followed by penile candidiasis, syphilitic chancres and highest for non-diseased images. Recall was lowest for penile cancer, which may be due to relative similarities in appearance of cancerous lesions with other disease states, or a lower overall number of images of cases of penile cancer on which the model was trained. Further model training using additional images from a more diverse sample is currently ongoing, which may further improve recall. Model precision was similarly highest for images of a non- diseased penis, while specificity was above 90% for all five diseases. Importantly, as this model was trained on images of these five disease classifications and non-diseased images, evaluation of the performance of the model among images of other penile diseases, such as other causes of genital ulcer disease (e.g., Haemophilus ducry infection), will be essential to validate model specificity.

Barriers to accessing care for sexual health, including sexually transmitted diseases, are numerous. In both high- and low-resource settings, fear of positive test results as well as social stigmatization and shame limit healthcare seeking.[27-29] In low- resource settings, individuals may not be able to reach a healthcare center,[29] and even among those who do, diagnostic options for such diseases are often limited or non- existent. Therefore, user-guided, mobile, image-based classification tools can facilitate rapid disease diagnoses in various contexts. Application of our model in settings where access to and utilization of sexual health services remains limited may increase case identification, facilitate earlier treatment, and reduce both the spread of disease and its consequent morbidity.

Such a user-accessible digital tool is predicated on the near-universal availability of smartphones, which, even in many low-resource areas, are ubiquitous; in 2022 there were an estimated 6.4 billion smartphone mobile network subscriptions worldwide.[30] However, some settings without smartphones or internet access will not be able to utilize this tool. Further, correctly establishing the diagnosis must be paired with patient education interventions and linkage to care to facilitate treatment. The HeHealth mobile platform has attempts to accomplish both by sending users that are classified as having one of the five diseases educational material specific to the diagnosis, such as common symptoms, what type of confirmatory testing is would be required, and treatment options. The app also provides links for additional resources to facilitate linkage to care. No outcome data, however, are available among HeHealth users.

Further assessment of the model also remains to be done. The performance of the model thus far has been based on images with a pre-existing classification. Comparisons with gold-standard diagnostics are warranted, such as a comparison of the diagnostic accuracy of the model with that of visual inspection by expert clinicians paired with microbiologic assessment where indicated (e.g., HSV polymerase chain reaction testing for herpes eruptions, fungal scraping for candidiasis, and serology as well as darkfield microscopy for Treponema pallidum infection). Additionally, follow-up data, if available could facilitate assessment of the clinical impact of the HeHealth platform. Further model building could seek to include additional disease classifications, such as lichen sclerosis or lichen planus, or seek to develop a similar model among vulvovaginal diseases. The possible applications of this technology are near boundless, and have the potential to address numerous health disparities.

## 4.1 Limitations

Our study had several limitations. First, the model was based on images with pre-assigned classifications by expert clinicians, not by images of diseases confirmed via microbiologic or histologic testing. Second, the performance of the model was assessed on a relatively small number of images, limiting the precision of our findings. Further evaluation on larger prospective datasets is warranted. However, given the potential applications of this tool, we feel those limitations do not negate the importance of our findings.

# 5 Conclusion

We report on the development of a machine-learning model for classifying five penile diseases. That model was developed using both clinical and augmented images, and demonstrated excellent performance on a validation dataset. Further assessment of model performance is warranted on larger, more diverse datasets. That model is currently in use globally and has the potential to improve access to diagnostic services for penile diseases.

# Figures and Tables

**Figure 1: Number and Type of Images Used for Training of the Machine-Learning Model Predicting Penile Pathology Classification**

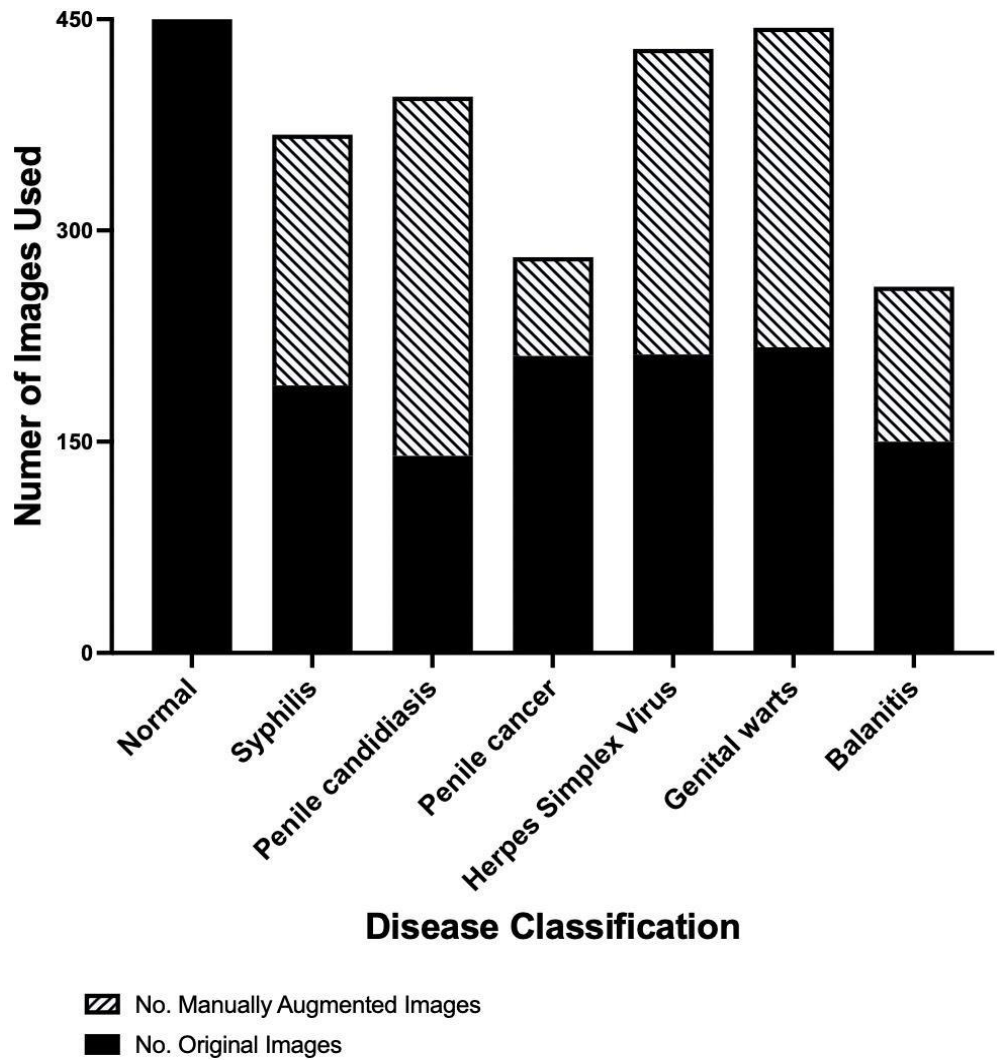

**Figure 1 Legend:** The figure shows the number of original and manually augmented images used across the five penile pathology classifications of interest (syphilitic chancres, penile candidiasis, penile cancer, herpes eruption, and genital warts).

**Table 1: Outcomes of the Machine-Learning Model for Visually Classifying Five Penile Diseases**

| | No. Images (n=239) | True Positive | False Positive | True Negative | False Negative | Recall or Sensitivity (95% CI) | | Precision (95% CI) | | Specificity (95% CI) | | F1-Score |
|---|---|---|---|---|---|---|---|---|---|---|---|---|
| Genital Warts | 45 | 43 | 7 | 187 | 2 | 0.956 | (0.849 - 0.995) | 0.860 | (0.764 - 0.956) | 0.964 | (0.927 - 0.985) | 0.909 |
| Herpes Eruption | 43 | 40 | 6 | 190 | 3 | 0.930 | (0.810 - 0.985) | 0.870 | (0.772 - 0.967) | 0.969 | (0.935 - 0.989) | 0.917 |
| Penile Cancer | 29 | 23 | 3 | 207 | 6 | 0.793 | (0.603 - 0.920) | 0.885 | (0.762 - 0.999) | 0.986 | (0.959 - 0.997) | 0.932 |
| Penile Candidiasis | 40 | 35 | 1 | 198 | 5 | 0.875 | (0.732 - 0.958) | 0.972 | (0.919 - 0.999) | 0.995 | (0.972 - 0.999) | 0.983 |
| Syphilis | 37 | 32 | 3 | 199 | 5 | 0.865 | (0.712 - 0.955) | 0.914 | (0.822 - 0.999) | 0.985 | (0.957 - 0.999) | 0.948 |
| Non-Diseased | 45 | 44 | 2 | 192 | 1 | 0.978 | (0.882 - 0.999) | 0.957 | (0.852 - 0.995) | 0.990 | (0.963 - 0.999) | 0.973 |

**Table 2: Distribution and Characteristics of Randomly Selected HeHealth Users Between July and October 2023**

| | Overall | | United States | | Singapore | | Canada | | United Kingdom | | Vietnam | |
|---|---|---|---|---|---|---|---|---|---|---|---|---|
| | No. | % | No. | % | No. | % | No. | % | No. | % | No. | % |
| Total | 437 | - | 271 | 62.0% | 64 | 14.6% | 41 | 9.4% | 40 | 9.2% | 21 | 4.8% |
| **Age (years)** | | | | | | | | | | | | |
| 18-30 | 277 | 63.4% | 176 | 64.9% | 37 | 57.8% | 26 | 63.4% | 25 | 62.5% | 13 | 61.9% |
| 31-50 | 144 | 33.0% | 87 | 32.1% | 23 | 35.9% | 14 | 34.1% | 13 | 32.5% | 7 | 33.3% |
| > 50 | 16 | 3.7% | 8 | 3.0% | 4 | 6.3% | 1 | 2.4% | 2 | 5.0% | 1 | 4.8% |
| **Symptoms** | | | | | | | | | | | | |
| Penile pain | 140 | 32.0% | 84 | 31.0% | 26 | 40.6% | 11 | 26.8% | 11 | 27.5% | 8 | 38.1% |
| Penile discharge | 110 | 25.2% | 70 | 25.8% | 12 | 18.8% | 10 | 24.4% | 11 | 27.5% | 7 | 33.3% |
| Pain/burning when urinating | 114 | 26.1% | 81 | 29.9% | 8 | 12.5% | 9 | 22.0% | 10 | 25.0% | 6 | 28.6% |
| None of the above / other | 242 | 55.4% | 149 | 55.0% | 38 | 59.4% | 24 | 58.5% | 21 | 52.5% | 10 | 47.6% |
| **Last sexual contact** | | | | | | | | | | | | |
| < 1 month ago | 248 | 56.8% | 169 | 62.4% | 30 | 46.9% | 20 | 48.8% | 21 | 52.5% | 8 | 38.1% |
| 1-3 months ago | 129 | 29.5% | 70 | 25.8% | 27 | 42.2% | 14 | 34.1% | 10 | 25.0% | 8 | 38.1% |
| > 3 months ago | 49 | 11.2% | 27 | 10.0% | 4 | 6.3% | 6 | 14.6% | 8 | 20.0% | 4 | 19.0% |
| Never had sex | 11 | 2.5% | 5 | 1.8% | 3 | 4.7% | 1 | 2.4% | 1 | 2.5% | 1 | 4.8% |

**Figure 2: De-Identified Images Submitted by HeHealth Users Between July and October 2023**

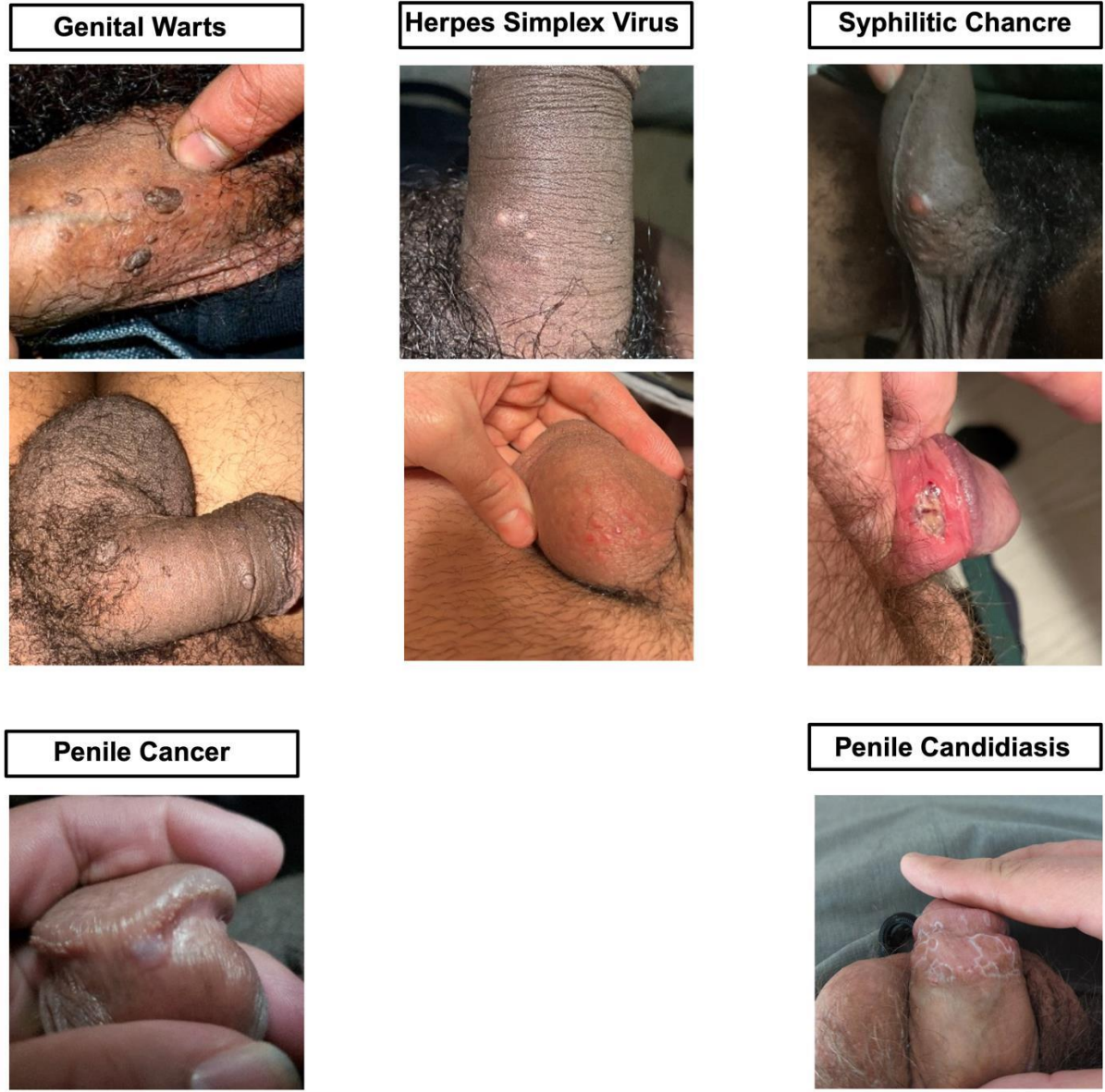

**Figure 2 Legend:** The figure shows images submitted by HeHealth users classified and classified into one of five penile disease categories by the model.